\documentclass{article}%
\usepackage{amsmath}
\usepackage{amsfonts}
\usepackage{amssymb}
\usepackage{graphicx}%
\providecommand{\U}[1]{\protect\rule{.1in}{.1in}}

\begin{document}

\title{Robust principal components for irregularly spaced longitudinal data}
\author{Ricardo A. Maronna\\University of La Plata and University of Buenos Aires (rmaronna@retina.ar)}
\date{\today}
\maketitle

\begin{abstract}
Consider longitudinal data $x_{ij},$ with $i=1,...,n$ and $j=1,...,p_{i},$
where $x_{ij}$ is the $j-$th observation of the random function $X_{i}\left(
.\right)  $ observed at time $t_{j}.$ The goal of this paper is to develop a
parsimonious representation of the data by a linear combination of a set of
$q$ smooth functions $H_{k}\left(  .\right)  $ ($k=1,..,q)$ in the sense that
$x_{ij}\approx\mu_{j}+\sum_{k=1}^{q}\beta_{ki}H_{k}\left(  t_{j}\right)  ,$
such that it fulfills three goals: it is\ resistant to atypical $X_{i}$'s
(\textquotedblleft case contamination\textquotedblright), it is resistant to
isolated gross errors at some $t_{ij}$ (\textquotedblright cell
contamination\textquotedblright), and it\ can be applied when some of the
$x_{ij}$ are missing (\textquotedblright irregularly spaced ---or
'incomplete'-- data\textquotedblright).

Two approaches will be proposed for this problem. One deals with the three
goals stated above, and is based on ideas similar to MM-estimation (Yohai
1987). The other is a simple and fast estimator which can be applied to
complete data with case- and cellwise contamination, and is based on applying
a standard robust principal components estimate and smoothing the principal
directions. Experiments with real and simulated data suggest that with
complete data the simple estimator outperforms its competitors, while the MM
estimator is competitive for incomplete data.


\end{abstract}

\section{Introduction}

Consider longitudinal data $x_{ij},$ with $i=1,...,n$ and $j=1,...,p,$ where
$x_{ij}$ is the $j-$th observation of the random function $X_{i}\left(
.\right)  $ observed at time $t_{j}.$ The goal of this paper is to develop a
parsimonious representation of the data by a linear combination of a set of
$q$ smooth functions $H_{k}\left(  .\right)  $ ($k=1,..,q)$ in the sense that
\begin{equation}
x_{ij}\approx\mu_{j}+\sum_{k=1}^{q}\beta_{ik}H_{k}\left(  t_{j}\right)  ,
\label{defxhat1}%
\end{equation}
such that it fulfills three goals: it is\ resistant to atypical $X_{i}$'s
(\textquotedblleft case contamination\textquotedblright), it is resistant to
isolated gross errors at some $t_{j}$ (\textquotedblright cell
contamination\textquotedblright), and it\ can be applied when some of the
$x_{ij}$ are missing (\textquotedblright irregularly spaced ---or
'incomplete'-- data\textquotedblright).

Among the abundant literature on this subject, Bali et al. (2011) propose an
approach based on projection pursuit; Boente et al. (2015) and Cevallos (2016)
transform the functional data to lower dimensional data through a spline basis
representation, to which they apply robust principal components based on the
minimization of a robust scale; and Lee et al. (2013) propose a penalized
M-estimator. These approaches hold only for complete data with casewise
contamination. James et al. (2001) and Yao et al (2005) propose estimators for
incomplete data. They require computing sample means and covariances and are
therefore not robust.

Two approaches will be proposed for this problem. One deals with the three
goals stated above, and is based on ideas similar to MM-estimation (Yohai
1987); the other is a very simple and fast estimator which can be applied to
complete data with case- and cellwise contamination. It is based on applying a
standard robust principal components estimate and smoothing the principal
directions, and will be called the \textquotedblleft Naive\textquotedblright\ estimator.

The contents of the paper are as follows. Sections \ref{secDefMM} and
\ref{secDefNaive} present the MM- and the Naive estimators. Sections
\ref{SecSimuComple} and \ref{SecComplExample} compare their performances for
complete data to those of former proposals through simulations and a real data
example. Sections \ref{SecIncomlExample} and \ref{secSimuIncomplete} compare
the performance of MM for incomplete data to that of a former proposal through
simulations and a real data example. Finally Sections \ref{SecCompMM} and
\ref{secNaifDetails} give the details of the computation of the MM- and Naive estimates.

\section{The \textquotedblleft MM\textquotedblright\ estimator\label{secDefMM}%
}

Let $B_{l}\left(  t\right)  $ ($l=1,...,m)$ be a basis of B-splines defined on
an interval containing $t_{1},...,t_{p}.$ Call $q$ the desired number of components.

For given $\boldsymbol{\alpha}\mathbf{=}[\alpha_{kl}]$, $\boldsymbol{\beta
}\mathbf{=[}\beta_{ik}\mathbf{]}$ and $\boldsymbol{\mu}=[\mu_{j}],$ with
$i=1,..,n,$ $j=1,...,p,$ $l=1,..,m$ and $k=1,...,q$ define
\[
\widehat{x}_{ij}\left(  \boldsymbol{\alpha.\beta,\mu}\right)  =\mu_{j}%
+\sum_{k=1}^{q}\beta_{ik}H_{k}\left(  t_{j}\right)  \
\]
with%
\begin{equation}
H_{k}\left(  t\right)  =\sum_{l=1}^{m}\alpha_{kl}B_{l}\left(  t\right)  .
\label{defH}%
\end{equation}

Then the proposed estimator is given by%
\begin{equation}
\left(  \widehat{\boldsymbol{a}},\widehat{\boldsymbol{\beta}},\widehat
{\boldsymbol{\mu}}\right)  =\arg\min_{\boldsymbol{\alpha,\beta,\mu}}\sum
_{i=1}^{n}\sum_{j\in J_{i}}\widehat{\sigma}_{j}^{2}\rho\left(  \frac
{x_{ij}-\widehat{x}_{ij}\left(  \boldsymbol{\alpha,\beta,\mu}\right)
}{\widehat{\sigma}_{j}}\right)  . \label{defineMM}%
\end{equation}
where $\widehat{\sigma}_{j}$ are previously computed local scales and
$J_{i}=\{j:x_{ij}$\ is non-missing$\}.$ If $\rho\left(  t\right)  =t^{2}$ and
the data are complete, the estimator coincides with the classical principal
components (PC) estimator. The matrix of principal directions
(\textquotedblleft eigenvectors\textquotedblright) is%
\begin{equation}
\mathbf{E=B}\widehat{\mathbf{\alpha}}^{\prime}\in R^{p\times q},
\label{defEmatrix}%
\end{equation}
where $\mathbf{B}$ is the matrix with elements $b_{jl}=B_{l}\left(
t_{j}\right)  .$

This estimator will henceforth be called \textquotedblleft
MM-estimator\textquotedblright. It is computed iteratively, starting from a
deterministic initial estimator. Since the whole procedure is complex, the
details are postponed to Section \ref{SecCompMM}.

A robust measure of \textquotedblleft unexplained variance\textquotedblright%
\ for $q$ components is defined as%
\[
V_{q}=\sum_{i=1}^{n}\sum_{j\in J_{i}}\widehat{\sigma}_{j}^{2}\rho\left(
\frac{x_{ij}-\widehat{x}_{ij}\left(  \widehat{\boldsymbol{\alpha}%
}\boldsymbol{,}\widehat{\boldsymbol{\beta}}\boldsymbol{,}\widehat
{\boldsymbol{\mu}}\right)  }{\widehat{\sigma}_{j}}\right)  ,
\]
and the \textquotedblleft proportion of explained variance\textquotedblright%
\ is given by
\begin{equation}
u_{q}=1-\frac{V_{q}}{V_{0}}, \label{defPropExMM}%
\end{equation}
where $V_{0}$ is obtained from (\ref{defineMM}) with $\mathbf{\alpha=0}$ and
$\mathbf{\beta=0.}$

The number of spline knots is of the form $[p/K],$ where $[.]$ denotes the
integer part. Choosing $K$ through cross-validation did not yield better
results than using a fixed value. After a series of exploratory simulations it
was decided that $K=6$ yielded satisfactory results.

\section{The \textquotedblleft naive\textquotedblright%
\ estimator\label{secDefNaive}}

A simple proposal complete data is introduced. It consists of four steps:

\begin{enumerate}
\item \textquotedblleft Clean\textquotedblright\ local outliers by applying a
robust smoother to each $\mathbf{x}_{i}$ and then imputing atypical values.

\item Compute $q$ ordinary robust PC's of the \textquotedblleft
cleaned\textquotedblright\ $\mathbf{x}_{i}$s.

\item Smooth the $q$ eigenvectors using a B-spline basis.

\item Orthogonalize them.
\end{enumerate}

A robust \textquotedblleft proportion of unexplained
variance\textquotedblright\ can be computed as explained in Section
\ref{secNaifDetails}.

In order to simplify the exposition, the details are given in Section
\ref{secNaifDetails}.

\section{Complete data: simulation\label{SecSimuComple}}

Instead of using arbitrary functions for the simulation scenarios, it was
considered more realistic to take them from a real data set. We chose the Low
Resolution Spectroscopy (LRS) data set (Bay 1999), which contains $n=531$
spectra, on the\textquotedblleft red\textquotedblright\ and \textquotedblleft
blue\textquotedblright\ ranges. We deal with the \textquotedblleft
blue\textquotedblright\ data, which has $p=44.$

The first two classical components account respectively for 60\% and 33\% of
the variability. It was therefore considered that $q=2$ components was a good choice.

\begin{figure}
[h]
\begin{center}
\includegraphics[
height=8.8947cm,
width=8.8947cm
]%
{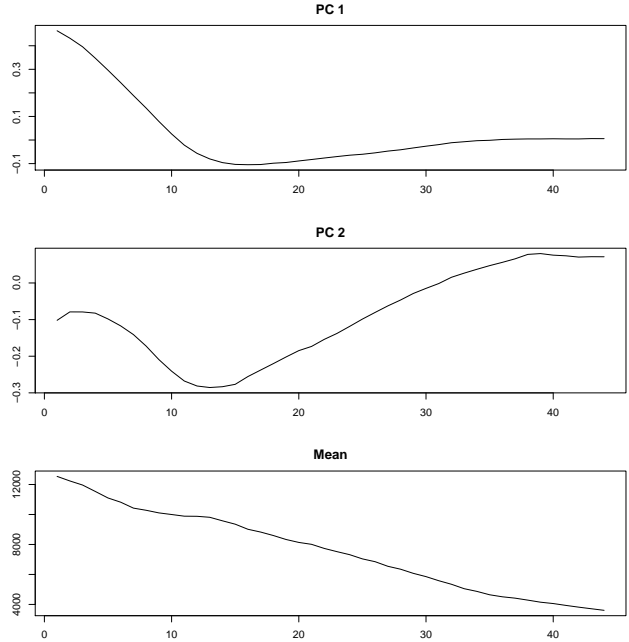}%
\caption{LRS Blue data:\ first two eigenvectors and mean vector.}%
\label{figTruePCs-LRS}%
\end{center}
\end{figure}

The first four classical eigenvectors $\mathbf{e}_{k}$ ($k=1,..,4)$ and the
mean vector $\boldsymbol{\mu}$ were computed, and a polynomial approximation
was fitted to them, so that they can be obtained for any $p.$ The first two
$\mathbf{e}_{k}$s and the mean vector are plotted in Figure
\ref{figTruePCs-LRS}%

For given $p$ and $q$ the data are generated as multivariate normal with mean
$\boldsymbol{\mu}$ and covariance matrix
\begin{equation}
\boldsymbol{\Sigma}\mathbf{=}\left[  \sigma_{ij}\right]  =\sum_{k=1}^{q}%
\pi_{k}\mathbf{e}_{k}\mathbf{e}_{k}^{\prime}+\pi_{0}\mathbf{I,}
\label{defSigma}%
\end{equation}
where now $\mathbf{\mu}$ and $\mathbf{e}_{k}$ are the original vectors
\textquotedblleft scaled\textquotedblright\ to the interval $[1,p]$ rather
than the original $[1,43],$ and $\mathbf{I}$ is the $p$-dimensional identity
matrix. The last term induces noise. The simulations were run for $q=2$ and 3
and under different configurations of the $\pi^{\prime}$s$.$

\subsection{Contamination}

Two types of contamination are considered. In \emph{case-wise} contamination,
a proportion $\varepsilon_{\mathrm{case}}$ of cases are replaced by
$\mathbf{x}_{0}=K\sqrt{\lambda_{1}}\mathbf{c,}$where $\mathbf{c}$ is
orthogonal to $\mathbf{e}_{k},$ $k=1,..,q,$ and $\lambda_{1}$ is the first
eigenvalue of $\mathbf{\Sigma.}$ The factor $\sqrt{\lambda_{1}}$ simplifies
the choice of the range of $K.$ Two choices for $\mathbf{c}$ were employed:
one was simply $\mathbf{e}_{q+1},$ the $(q+1)$-th eigenvector of
$\mathbf{\Sigma,}$ and the other was a random direction orthogonal to
$\mathbf{e}_{k},$ $k=1,..,q.$ Since the qualitative results yielded by both
options were similar, only those corresponding to the first one are shown.

The outlier size $K$ ranges between 0.10 and 3$,$ in order to find the
\textquotedblleft most malign\textquotedblright\ configurations for each estimator.

\begin{figure}
[h]
\begin{center}
\includegraphics[
height=8.8947cm,
width=8.8947cm
]%
{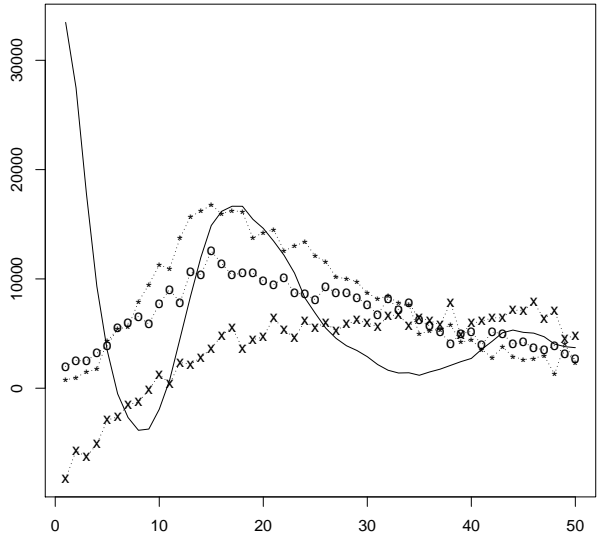}%
\caption{Simulation: three \textquotedblleft typical\textquotedblright\ cases
and one case outlier (--)}%
\label{figTipi1}%
\end{center}
\end{figure}

Figure \ref{figTipi1} shows for $p=50$ and $q=2$ three \textquotedblleft
typical\textquotedblright\ cases and a case outlier with $K=1.2$.%

In \emph{cell-wise} contamination, each cell is contaminated at random with
probability $\varepsilon_{\mathrm{cell}}=1/p.$ This means that each case has
in average one outlier, but about 63\% of the \emph{cases} have some outliers.
In contaminated cells, $x_{ij}$ is replaced by $x_{ij}+K\sigma_{jj}.$ Here $K$
ranges between 1 and 7.

\subsection{Evaluation}

Two issues have to be considered to evaluate an estimator's performance at a
given simulation scenario. The first is the precision with which the PCs
reconstruct the data, which means measuring the differences between the fitted
and the observed values. It was observed that even for \textquotedblleft
good\textquotedblright\ data, the distribution of the prediction errors
$x_{ij}-\widehat{x}_{ij}$ may have heavy tails. For this reason the fit of
each estimator was evaluated through its prediction Mean Absolute Error (MAE),
which was considered as less sensitive than the Mean Squared Error to some
atypical values.

For a given estimator call $\widehat{x}_{ij}$ the fitted values. For
\emph{case-wise} contamination call $I$ the set of non-contaminated rows. Then
define for a data set $[x_{ij}]$
\[
\mathrm{MAE=ave}\left\{  |x_{ij}-\widehat{x}_{ij}|:i\in I,\ j=1,..,p\right\}
.
\]
For \emph{cell-wise} contamination the average is taken over \emph{all }cells:%
\[
\mathrm{MAE=ave}\left\{  |x_{ij}-\widehat{x}_{ij}%
|:i=1,..,n,\ j=1,..,p\right\}  ,
\]
the justification being that the estimator should be able to \textquotedblleft
guess\textquotedblright\ the true value at the contaminated cell because of
the assumed underlying smoothness of the data.

Finally, for each estimator the average MAE over all Monte Carlo replications
is computed.

The second issue is the estimation of the \textquotedblleft
true\textquotedblright\ PCs. To this end, the angle $\alpha$ between the
estimated subspace and the subspace generated by $\{\mathbf{e}_{k},k=1,..,q\}$
was computed. Since it was thought that $\sin\left(  \alpha\right)  $ would be
easier to interpret (because it ranges between 0 and one), the averages of
$\sin\left(  \alpha\right)  $ are reported.

\section{Simulation results}

The estimators involved in the simulation are: Classic, \textquotedblleft
Naive\textquotedblright, MM, the S-estimator of Boente and Salibian-Barrera
(2015) using a B-spline basis, and the LTS estimator of Cevallos (2016). The R
code for the S and LTS estimators was kindly supplied by the authors.
Unfortunately it was not possible to obtain the code for the estimator of Lee
et al. (2013).

The dimension was set at $p=50.$ The number of components was chosen as $q=$ 2
and 3. Different configurations for the $\pi_{j}$s in (\ref{defSigma}) were
employed. The results given here are for $q=2$ and $\pi_{1}=0.6,$ $\pi
_{2}=0.3,$ $\pi_{3}=0.05.$ Since the qualitative results yielded by $q=3$ and
by other configurations of the $\pi_{j}$s are similar, they are not reported
here. The number of replications is 200. It would probably be more realistic
to fix a predetermined proportion of \textquotedblleft explained
variance\textquotedblright\ (say, 90\%) and let each estimator choose $q$
accordingly. While this can be done with the Classic, Naive and MM estimators,
it cannot be done with the S and LTS estimators.

\textquotedblleft Efficiency\textquotedblright\ will be loosely defined as
\textquotedblleft similitude with the Classical estimator when $\varepsilon
_{\mathrm{case}}=\varepsilon_{\mathrm{cell}}=0$\textquotedblright.

Table \ref{TabL1} shows the maximum averages (over $K$) of the MAEs for $p=50$
and two components.

\begin{center}%
\begin{table}[tbp] \centering
\begin{tabular}
[c]{cccccccc}\hline
$n$ & $\varepsilon_{\mathrm{case}}$ & $\varepsilon_{\mathrm{cell}}$ & Class. &
Naive & MM & S & LTS\\\hline
\multicolumn{1}{r}{100} & \multicolumn{1}{r}{0} & \multicolumn{1}{r}{0} &
\multicolumn{1}{r}{557} & \multicolumn{1}{r}{563} & \multicolumn{1}{r}{595} &
\multicolumn{1}{r}{563} & \multicolumn{1}{r}{600}\\
\multicolumn{1}{r}{} & \multicolumn{1}{r}{0.1} & \multicolumn{1}{r}{0} &
\multicolumn{1}{r}{2663} & \multicolumn{1}{r}{565} & \multicolumn{1}{r}{675} &
\multicolumn{1}{r}{564} & \multicolumn{1}{r}{638}\\
\multicolumn{1}{r}{} & \multicolumn{1}{r}{0.2} & \multicolumn{1}{r}{0} &
\multicolumn{1}{r}{2633} & \multicolumn{1}{r}{604} & \multicolumn{1}{r}{1051}
& \multicolumn{1}{r}{591} & \multicolumn{1}{r}{2708}\\
\multicolumn{1}{r}{} & \multicolumn{1}{r}{0} & \multicolumn{1}{r}{0.02} &
\multicolumn{1}{r}{2701} & \multicolumn{1}{r}{585} & \multicolumn{1}{r}{617} &
\multicolumn{1}{r}{1567} & \multicolumn{1}{r}{1593}\\\hline
\multicolumn{1}{r}{200} & \multicolumn{1}{r}{0} & \multicolumn{1}{r}{0} &
\multicolumn{1}{r}{282} & \multicolumn{1}{r}{287} & \multicolumn{1}{r}{298} &
\multicolumn{1}{r}{285} & \multicolumn{1}{r}{411}\\
\multicolumn{1}{r}{} & \multicolumn{1}{r}{0.1} & \multicolumn{1}{r}{0} &
\multicolumn{1}{r}{1334} & \multicolumn{1}{r}{288} & \multicolumn{1}{r}{351} &
\multicolumn{1}{r}{287} & \multicolumn{1}{r}{414}\\
\multicolumn{1}{r}{} & \multicolumn{1}{r}{0.2} & \multicolumn{1}{r}{0} &
\multicolumn{1}{r}{1323} & \multicolumn{1}{r}{334} & \multicolumn{1}{r}{547} &
\multicolumn{1}{r}{287} & \multicolumn{1}{r}{1406}\\
\multicolumn{1}{r}{} & \multicolumn{1}{r}{0} & \multicolumn{1}{r}{0.02} &
\multicolumn{1}{r}{892} & \multicolumn{1}{r}{288} & \multicolumn{1}{r}{295} &
\multicolumn{1}{r}{664} & \multicolumn{1}{r}{764}\\\hline
\end{tabular}
\caption{Simulation: maximum MAEs of estimators}\label{TabL1}%
\end{table}%

\end{center}

Table \ref{TabAngu} shows the maximum mean $\sin\left(  \alpha\right)  $ for
the same scenarios.

\begin{center}%
\begin{table}[tbp] \centering
\begin{tabular}
[c]{cccccccc}\hline
$n$ & $\varepsilon_{\mathrm{case}}$ & $\varepsilon_{\mathrm{cell}}$ & Class. &
Naive & MM & S & LTS\\\hline
\multicolumn{1}{r}{50} & \multicolumn{1}{r}{0} & \multicolumn{1}{r}{0} &
\multicolumn{1}{r}{0.014} & \multicolumn{1}{r}{0.015} &
\multicolumn{1}{r}{0.025} & \multicolumn{1}{r}{0.018} &
\multicolumn{1}{r}{0.055}\\
\multicolumn{1}{r}{} & \multicolumn{1}{r}{0.1} & \multicolumn{1}{r}{0} &
\multicolumn{1}{r}{0.998} & \multicolumn{1}{r}{0.016} &
\multicolumn{1}{r}{0.079} & \multicolumn{1}{r}{0.020} &
\multicolumn{1}{r}{0.070}\\
\multicolumn{1}{r}{} & \multicolumn{1}{r}{0.2} & \multicolumn{1}{r}{0} &
\multicolumn{1}{r}{0.999} & \multicolumn{1}{r}{0.036} &
\multicolumn{1}{r}{0.194} & \multicolumn{1}{r}{0.042} &
\multicolumn{1}{r}{0.878}\\
\multicolumn{1}{r}{} & \multicolumn{1}{r}{0} & \multicolumn{1}{r}{0.02} &
\multicolumn{1}{r}{0.619} & \multicolumn{1}{r}{0.016} &
\multicolumn{1}{r}{0.039} & \multicolumn{1}{r}{0.021} &
\multicolumn{1}{r}{0.058}\\\hline
\multicolumn{1}{r}{200} & \multicolumn{1}{r}{0} & \multicolumn{1}{r}{0} &
\multicolumn{1}{r}{0.011} & \multicolumn{1}{r}{0.012} &
\multicolumn{1}{r}{0.017} & \multicolumn{1}{r}{0.016} &
\multicolumn{1}{r}{0.210}\\
\multicolumn{1}{r}{} & \multicolumn{1}{r}{0.1} & \multicolumn{1}{r}{0} &
\multicolumn{1}{r}{0.998} & \multicolumn{1}{r}{0.013} &
\multicolumn{1}{r}{0.085} & \multicolumn{1}{r}{0.019} &
\multicolumn{1}{r}{0.211}\\
\multicolumn{1}{r}{} & \multicolumn{1}{r}{0.2} & \multicolumn{1}{r}{0} &
\multicolumn{1}{r}{0.999} & \multicolumn{1}{r}{0.059} &
\multicolumn{1}{r}{0.190} & \multicolumn{1}{r}{0.020} &
\multicolumn{1}{r}{0.879}\\
& 0 & 0.02 & 0.287 & 0.012 & 0.020 & 0.019 & 0.211\\\hline
\end{tabular}
\caption{Simulation: maximum  sin($\alpha$) of estimators}\label{TabAngu}%
\end{table}%

\end{center}

\subsection{Discussion}

Comparing the maximum MAEs it can be concluded that

\begin{itemize}
\item The Naive, S and MM estimators appear as reasonably efficient

\item For case-wise contamination, the Naive and S estimators appear as the
best, the latter being slightly better; LTS and MM have similar acceptable
performances for $\varepsilon_{\mathrm{case}}=0.1;$ when $\varepsilon
_{\mathrm{case}}=0.2,$ MM and especially LTS have poorer performances

\item For cell-wise contamination, the Naive and MM estimators are practically
unaffected by outliers, while the other estimators are clearly affected by
them.\bigskip
\end{itemize}

The results of Table \ref{TabAngu} for $\varepsilon_{\mathrm{cell}}=0$
parallel those of Table \ref{TabL1}. However, comparing the results for
$\varepsilon_{\mathrm{cell}}=0.02$ with those for $\varepsilon_{\mathrm{cell}%
}=\varepsilon_{\mathrm{case}}=0$ it is seen that S and LTS are almost
unaffected by cell outliers!. The seeming contradiction between these results
and those of Table \ref{TabL1} is explained by the fact that in
(\ref{defxhat1}), even if the $H_{k}$s are correctly estimated, the $\beta$s
must be robustly estimated too, and it is here that these two estimators fail.

\bigskip Examination of intermediate results reveals that the
\textquotedblleft weak spot\textquotedblright\ of MM lies at the starting
values, which require the estimation of a covariance matrix, which has to be
deterministic, fast, resistant to cell-wise outliers and computable for
incomplete data. The price for so many requirements is a decrease in
robustness for large $\varepsilon_{\mathrm{case}}.$ See Section
\ref{SecCompMM} for details.

\section{Complete data: a real example\label{SecComplExample}}

The chosen data set is the \textquotedblleft red\textquotedblright\ range of
the LRS data (Bay 1999), which has $n=531$ spectra with $p=49.$

The estimated proportions of explained variance with two components are

\begin{center}%
\begin{tabular}
[c]{rrr}\hline
Classic & Naive & MM\\
0.91 & 0.85 & 0.88\\\hline
\end{tabular}

\end{center}

It was therefore considered reasonable to choose $q=2$ components.

Since our goal is assessing how well do the fitted values approximate the data
(rather than outlier detection), estimates will be evaluated by their MAEs.
For a given estimate let the MAE of row $i=1,..,n$ be
\begin{equation}
m_{i}=\mathrm{ave}\left\{  \left\vert x_{ij}-\widehat{x}_{ij}\right\vert
:j=1,...,p\right\}  ,\ \label{der-m_i}%
\end{equation}
\qquad\qquad

The overall MAEs (averages of the $m_{i})$ are%
\[%
\begin{tabular}
[c]{ccccc}\hline
Class. & Naive & MM & S & LTS\\\hline
\multicolumn{1}{r}{184} & \multicolumn{1}{r}{160} & \multicolumn{1}{r}{162} &
\multicolumn{1}{r}{160} & \multicolumn{1}{r}{384}\\\hline
\end{tabular}
\ \ \ .
\]

\begin{figure}
[h]
\begin{center}
\includegraphics[
height=8.8947cm,
width=8.8947cm
]%
{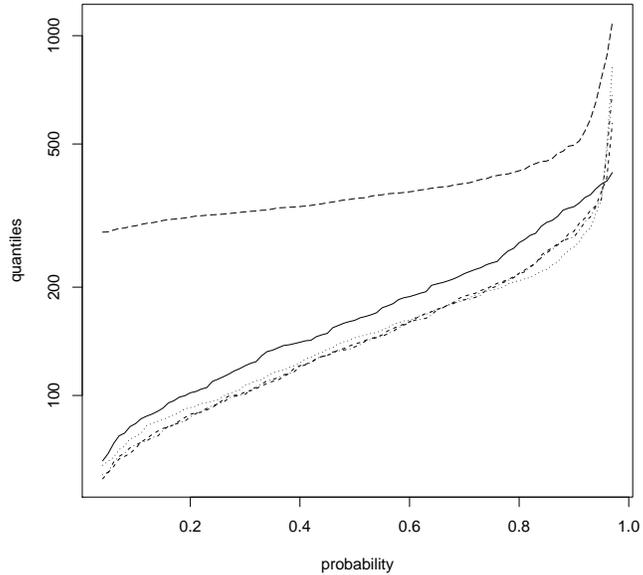}%
\caption{LRS data: quantiles of estimates' MAEs: Classical (---), Naive(- -),
MM (...), S (-.-.) and LTS (-- --)}%
\label{figLRSQuant_L1}%
\end{center}
\end{figure}

Figure \ref{figLRSQuant_L1} shows the quantiles (in log scale) of the MAEs
corresponding to the different estimates.%

\begin{figure}
[h]
\begin{center}
\includegraphics[
height=8.8947cm,
width=8.8947cm
]%
{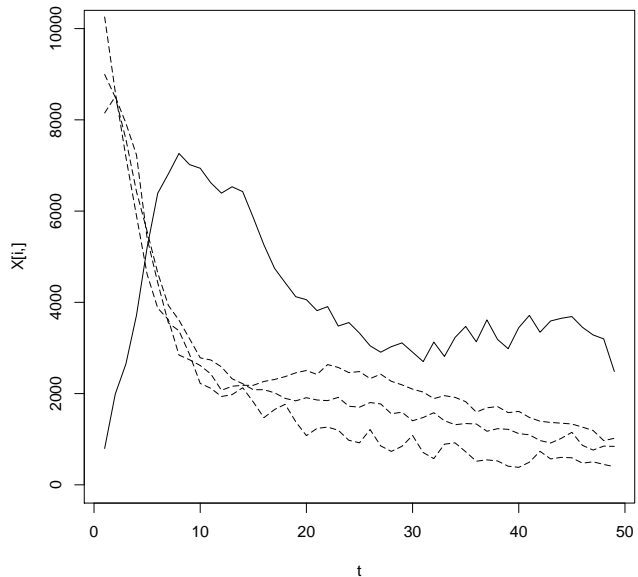}%
\caption{LRS data: three \textquotedblleft typical\textquotedblright\ cases
(dashed line) and an outlier (full line).}%
\label{FigOL_LRS}%
\end{center}
\end{figure}

The MM, Naive and S estimate give similar results; the classical estimate
gives slightly poorer results, and LTS shows a poor fit. The quantiles of the
classical estimate are larger than those of MM, Naive and S, except for the
top 4\%. This fact can be attributed to outlying cases. Figure \ref{FigOL_LRS}
shows three \textquotedblleft typical\textquotedblright\ cases corresponding
to the quartiles of the Naive estimate's $m_{i}$ (dashed lines) and to the
0.96-quantile (solid line). The latter appears as totally atypical. One may
thus interpret that the classical estimate attempts to fit all cases,
including the atypical ones, at the expense of the typical ones, while the
robust estimates downweight the latter.%

\begin{figure}
[h]
\begin{center}
\includegraphics[
height=8.8989cm,
width=14.5145cm
]%
{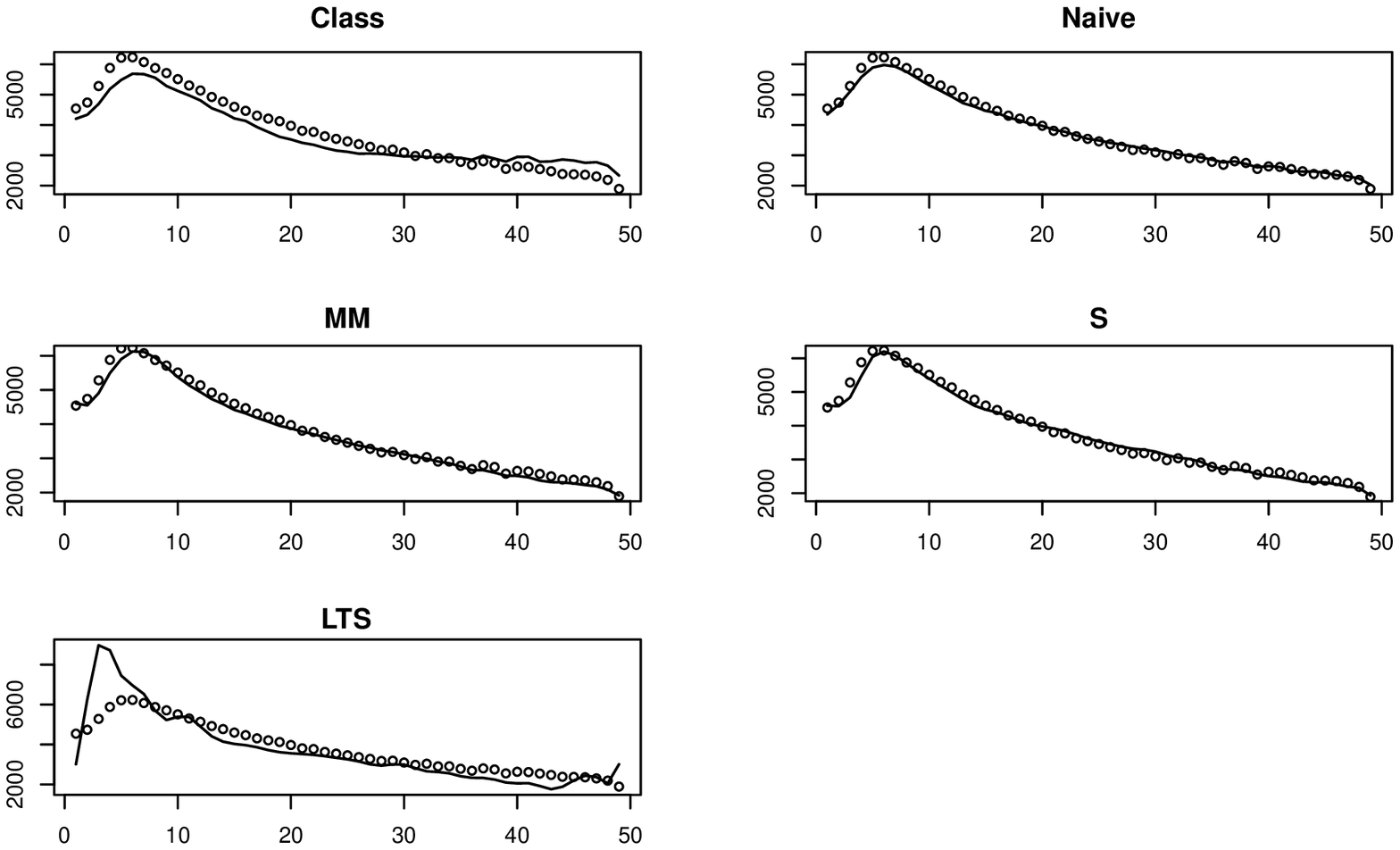}%
\caption{LRS data: case 338 (o) and fits by different estimates (---)}%
\label{Fig-LRS_338}%
\end{center}
\end{figure}

For a closer look at the behavior of LTS, Figure \ref{Fig-LRS_338} shows the
fit to case 338, a typical one. It is seen that the Naive, S and MM estimators
give good fits and the Classical performs a bit worse, while LTS gives wrong
values at the extremes. This behavior occurs in most cases of this data set.%

\section{Computing times}

The next table shows the average computing times in seconds for $n=4p$ and
number of components $q,$ on a PC with a 3.60 GH processor with 4 cores. Here
\textquotedblleft S\textquotedblright\ denotes the parallelized version of the
S estimator's code.

\begin{center}%
\begin{table}[tbp] \centering
\begin{tabular}
[c]{ccccccc}\hline
\multicolumn{1}{|c}{$p$} & $n$ & $q$ & Naive & MM & LTS & S\\
\multicolumn{1}{r}{50} & \multicolumn{1}{r}{200} & \multicolumn{1}{r}{2} &
\multicolumn{1}{r}{0.07} & \multicolumn{1}{r}{2.65} &
\multicolumn{1}{r}{11.80} & 22.91\\
\multicolumn{1}{r}{} & \multicolumn{1}{r}{} & \multicolumn{1}{r}{4} &
\multicolumn{1}{r}{0.07} & \multicolumn{1}{r}{3.66} &
\multicolumn{1}{r}{12.69} & 23.13\\
\multicolumn{1}{r}{100} & \multicolumn{1}{r}{400} & \multicolumn{1}{r}{2} &
\multicolumn{1}{r}{0.25} & \multicolumn{1}{r}{14.75} &
\multicolumn{1}{r}{19.18} & 103.80\\
\multicolumn{1}{r}{} & \multicolumn{1}{r}{} & \multicolumn{1}{r}{4} &
\multicolumn{1}{r}{0.25} & \multicolumn{1}{r}{22.24} &
\multicolumn{1}{r}{21.98} & 107.37\\
\multicolumn{1}{r}{200} & \multicolumn{1}{r}{800} & \multicolumn{1}{r}{2} &
\multicolumn{1}{r}{1.20} & \multicolumn{1}{r}{105.73} &
\multicolumn{1}{r}{32.10} & 710.31\\
\multicolumn{1}{r}{} & \multicolumn{1}{r}{} & \multicolumn{1}{r}{4} &
\multicolumn{1}{r}{1.21} & \multicolumn{1}{r}{157.84} &
\multicolumn{1}{r}{38.60} & 742.25\\\hline
\end{tabular}
\caption{Computing times of  estimators}\label{TabTimes}%
\end{table}%

\end{center}

\bigskip It is seen that the Naive estimator is by far the fastest, and S is
the slowest; MM is faster than LTS for $p\leq100,$ but slower otherwise. The
computing times of S and LTS are unaffected by $q;$ the computing time of MM
increases with $q$ (because components are computed one at a time)

As a final conclusion: considering efficiency, robustness and speed, it may be
concluded that for complete data, the Naive estimator is to be recommended.

\section{Incomplete data: a real example\label{SecIncomlExample}}

We begin our treatment of incomplete data by introducing a new data set, which
belongs to the Multicenter AIDS Cohort Study (MACS) and is available at
https://statepi.jhsph.edu/macs/pdt.html. It contains $n=305$ cases, with
$p=48.$ The number of observations per case varies between 1 and 9. There is a
total of 2001 measurements, which yields a \textquotedblleft decimation
rate\textquotedblright%
\begin{equation}
d=\frac{\#(\mathrm{measurements})}{np}=0.137. \label{decima}%
\end{equation}

The MM-estimator was compared to the procedure \textquotedblleft Principal
Analysis by Conditional Estimation\textquotedblright\ (PACE) (Yao et al 2005),
which is implemented in the package \texttt{fdapace}\textbf{ }available at the
authors' web site.

The proportions of explained variance for one and two components were 0.92 and
0.98 for PACE, and 0.77 and 0.81 for MM. It was decided to take $q=2.$

\begin{figure}
[h]
\begin{center}
\includegraphics[
height=8.8947cm,
width=8.8947cm
]%
{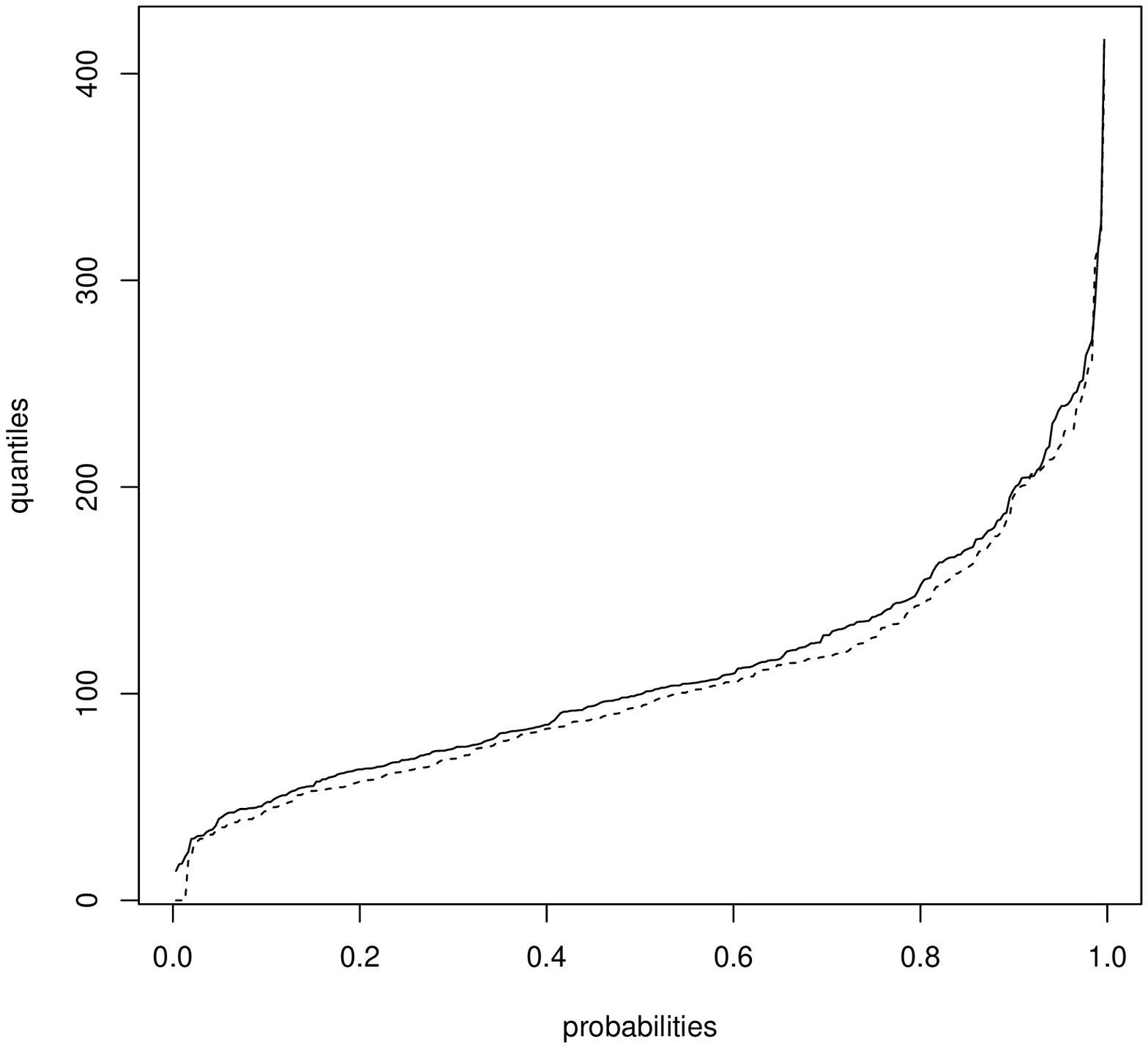}%
\caption{MACS data: case MAEs of PACE(---) and MM ( -)}%
\label{FigCompaMACs}%
\end{center}
\end{figure}

The overall MAEs of PACE and MM are respectively 93.2 and 87.5. Figure
\ref{FigCompaMACs} compares the quantiles of the MAEs $m_{i}$ (\ref{der-m_i})
from both estimates. It is seen that their performances are similar, with MM
doing slightly better.%

The proportions of explained variance from MM are much lower than those from
PACE. The reason is that they employ different measures of variability.
Outliers in the first principal axes inflate the ordinary variance but
contribute little to a robust variance. A similar phenomenon is observed in
(Maronna et al, 2006, ps. 213-14).

\section{Incomplete data: simulation: \label{secSimuIncomplete}}

Two scenarios were chosen for the simulations. One is the same employed in
Section \ref{SecSimuComple}. The other is based on the MACS data set; the mean
vector $\mathbf{\mu}$ and the first principal directions $\mathbf{e}_{k}$ of
the data were estimated with PACE, and then employed in the same way as with
the LRS data. They will be called \textquotedblleft LRS
scenario\textquotedblright\ and \textquotedblleft MACS
scenario\textquotedblright.

For each scenario a complete data set was generated, and then each cell was
kept at random with probability $d,$ the \textquotedblleft decimation
rate\textquotedblright\ as in (\ref{decima}), and the others set as
\textquotedblleft missing\textquotedblright. The values of $d$ were 0.25 and
0.5. The outlier size $K$ ranged between 1 and 6. The values of $\pi_{1},$
$\pi_{2}$ and $\pi_{3}$ are the same as in Section \ref{SecSimuComple}. The
dimension was set as $p=50,$ and the sample size $n$ took on the values 100
and 200. Since both values yielded similar results, only those corresponding
to the first one are shown.

Table \ref{TabMACNew} shows the results for the MACS scenario with $q=2$. It
is seen that MM is practically unaffected by the contamination. The MAEs of
PACE are heavily affected by both case- and cell-contamination; the angle is
also heavily affected by case contamination, but not so much by cell contamination.

\begin{center}%
\begin{table}[tbp] \centering
\begin{tabular}
[c]{ccccccc}\hline
$d$ & $\varepsilon_{\mathrm{case}}$ & $\varepsilon_{\mathrm{cell}}$ &
\multicolumn{2}{c}{MAE} & \multicolumn{2}{c}{Angle}\\\hline
\multicolumn{1}{l}{} & \multicolumn{1}{l}{} & \multicolumn{1}{l}{} & PACE &
MM & PACE & MM\\\cline{4-7}%
\multicolumn{1}{r}{0.5} & \multicolumn{1}{r}{0} & \multicolumn{1}{r}{0} &
\multicolumn{1}{r}{26.49} & \multicolumn{1}{r}{23.64} &
\multicolumn{1}{r}{0.066} & \multicolumn{1}{r}{0.058}\\
\multicolumn{1}{r}{} & \multicolumn{1}{r}{0.05} & \multicolumn{1}{r}{0} &
\multicolumn{1}{r}{155.71} & \multicolumn{1}{r}{24.61} &
\multicolumn{1}{r}{0.996} & \multicolumn{1}{r}{0.072}\\
\multicolumn{1}{r}{} & \multicolumn{1}{r}{0.1} & \multicolumn{1}{r}{0} &
\multicolumn{1}{r}{210.43} & \multicolumn{1}{r}{26.46} &
\multicolumn{1}{r}{0.999} & \multicolumn{1}{r}{0.096}\\
\multicolumn{1}{r}{} & \multicolumn{1}{r}{0} & \multicolumn{1}{r}{0.02} &
\multicolumn{1}{r}{55.40} & \multicolumn{1}{r}{23.72} &
\multicolumn{1}{r}{0.087} & \multicolumn{1}{r}{0.058}\\
\multicolumn{1}{r}{} & \multicolumn{1}{r}{0} & \multicolumn{1}{r}{0.05} &
\multicolumn{1}{r}{97.74} & \multicolumn{1}{r}{23.80} &
\multicolumn{1}{r}{0.112} & \multicolumn{1}{r}{0.058}\\\hline
\multicolumn{1}{r}{0.25} & \multicolumn{1}{r}{0} & \multicolumn{1}{r}{0} &
\multicolumn{1}{r}{28.97} & \multicolumn{1}{r}{25.27} &
\multicolumn{1}{r}{0.097} & \multicolumn{1}{r}{0.097}\\
\multicolumn{1}{r}{} & \multicolumn{1}{r}{0.05} & \multicolumn{1}{r}{0} &
\multicolumn{1}{r}{156.31} & \multicolumn{1}{r}{25.62} &
\multicolumn{1}{r}{0.987} & \multicolumn{1}{r}{0.100}\\
\multicolumn{1}{r}{} & \multicolumn{1}{r}{0.1} & \multicolumn{1}{r}{0} &
\multicolumn{1}{r}{209.98} & \multicolumn{1}{r}{27.02} &
\multicolumn{1}{r}{0.996} & \multicolumn{1}{r}{0.112}\\
\multicolumn{1}{r}{} & \multicolumn{1}{r}{0} & \multicolumn{1}{r}{0.02} &
\multicolumn{1}{r}{64.13} & \multicolumn{1}{r}{25.70} &
\multicolumn{1}{r}{0.132} & \multicolumn{1}{r}{0.097}\\
\multicolumn{1}{r}{} & \multicolumn{1}{r}{0} & \multicolumn{1}{r}{0.05} &
\multicolumn{1}{r}{111.27} & \multicolumn{1}{r}{27.46} &
\multicolumn{1}{r}{0.173} & \multicolumn{1}{r}{0.098}\\\hline
\end{tabular}
\caption{Simulation with MACS scenario: maximum MAEs and mean angles for  $p=50$ and  $n=100$}\label{TabMACNew}%
\end{table}%

\end{center}

\begin{figure}
[h]
\begin{center}
\includegraphics[
height=7.5788cm,
width=10.0925cm
]%
{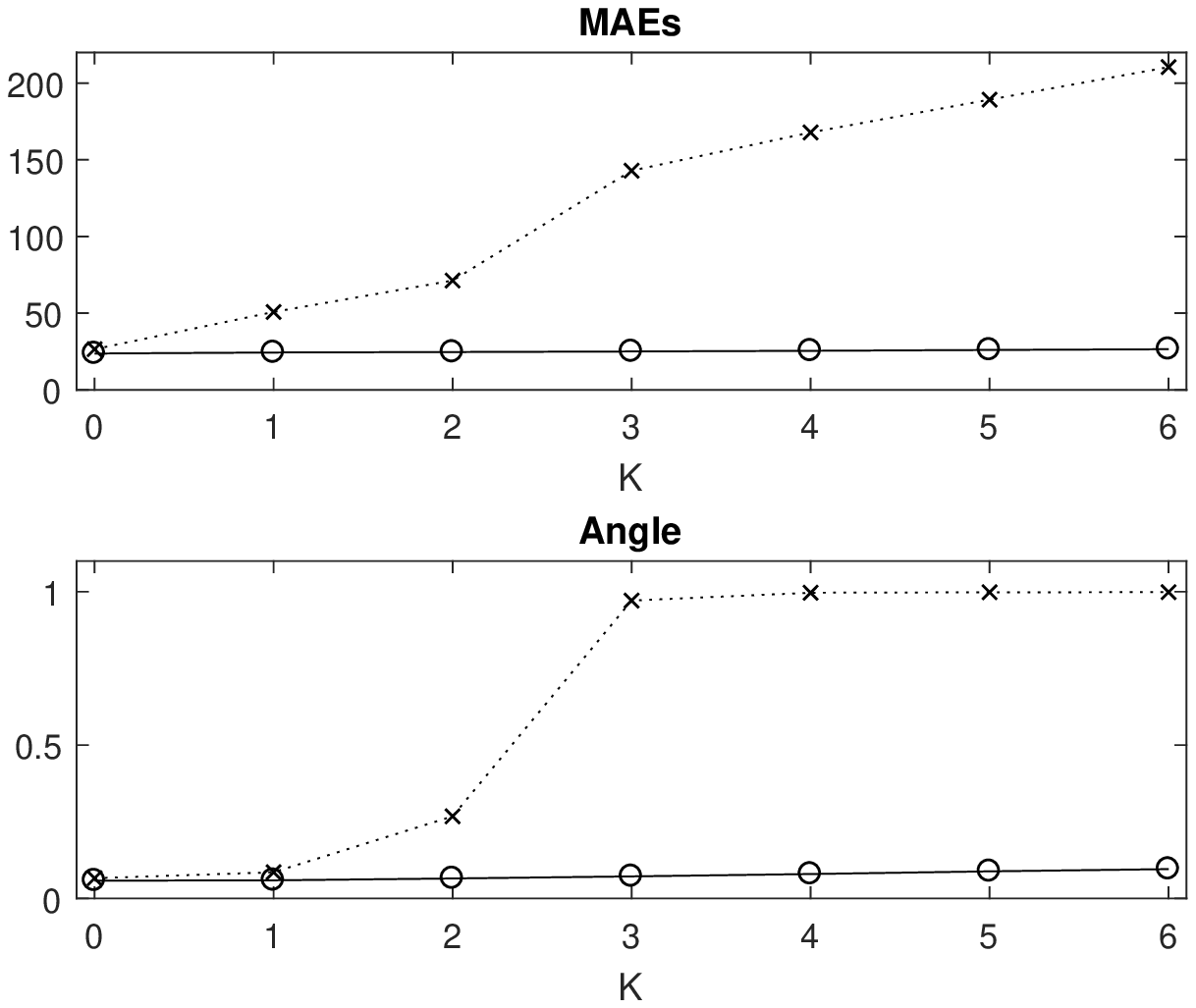}%
\caption{Simulation with MACS scenario: MAEs of PACE(x) and MM (o) as a
function of outlier size $K$ for $\varepsilon_{\mathrm{case}}=0.1$ and
$d=0.5$}%
\label{FigMAC_Case}%
\end{center}
\end{figure}

\begin{figure}
[h]
\begin{center}
\includegraphics[
height=8.047cm,
width=10.7209cm
]%
{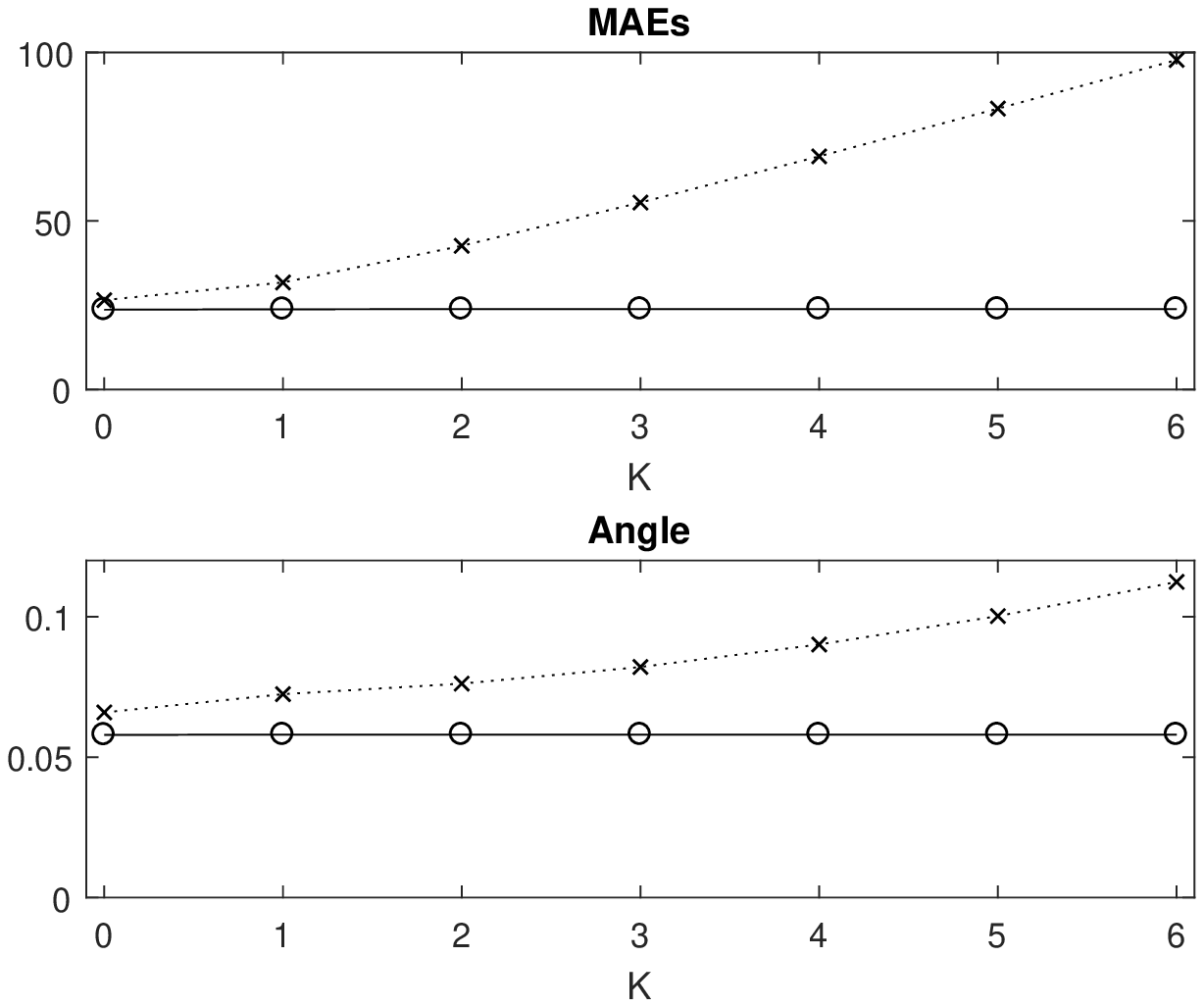}%
\caption{Simulation with MACS scenario: MAEs of PACE(x) and MM (o) as a
function of outlier size $K$ for $\varepsilon_{\mathrm{cell}}=0.05$ and
$d=0.5$}%
\label{FigMAC_Cell}%
\end{center}
\end{figure}

Figure \ref{FigMAC_Case} shows the MAEs and mean angles as a function of $K$
for $d=0.5.$ $\varepsilon_{\mathrm{case}}=0.1$ and $\varepsilon_{\mathrm{cell}%
}=0,$ and Figure \ref{FigMAC_Cell} does the same for $\varepsilon
_{\mathrm{case}}=0$ and $\varepsilon_{\mathrm{cell}}=0.05.$ The values for
$K=0$ are those corresponding $\varepsilon_{\mathrm{case}}=\varepsilon
_{\mathrm{cell}}=0.$ It is seen that in all cases MM remains practically
constant with $K;$ when $\varepsilon_{\mathrm{case}}=0.1$ PACE is clearly
affected when $K\geq2;$ and when $\varepsilon_{\mathrm{cell}}=0.05$ its MAEs
are also affected but the angle shows more resistance.%

%

A surprising feature of Table \ref{TabMACNew} is that for $\varepsilon
_{\mathrm{case}}=\varepsilon_{\mathrm{cell}}=0$ MM\ is slightly better than
PACE. Changing the random generator seed yielded similar results.

Table \ref{TabLRSCepiM} shows the results for the LRS scenario.

\begin{center}%
\begin{table}[tbp] \centering
\begin{tabular}
[c]{ccccccc}\hline
$d$ & $\varepsilon_{\mathrm{case}}$ & $\varepsilon_{\mathrm{cell}}$ &
\multicolumn{2}{c}{MAE} & \multicolumn{2}{c}{Angle}\\\hline
&  &  & PACE & MM & PACE & MM\\\cline{4-7}%
\multicolumn{1}{r}{0.5} & \multicolumn{1}{r}{0} & \multicolumn{1}{r}{0} &
\multicolumn{1}{r}{778} & \multicolumn{1}{r}{637} & \multicolumn{1}{r}{0.171}
& \multicolumn{1}{r}{0.082}\\
\multicolumn{1}{r}{} & \multicolumn{1}{r}{0.05} & \multicolumn{1}{r}{0} &
\multicolumn{1}{r}{2664} & \multicolumn{1}{r}{819} & \multicolumn{1}{r}{
0.644} & \multicolumn{1}{r}{0.086}\\
\multicolumn{1}{r}{} & \multicolumn{1}{r}{0.1} & \multicolumn{1}{r}{0} &
\multicolumn{1}{r}{3978} & \multicolumn{1}{r}{697} & \multicolumn{1}{r}{0.968}
& \multicolumn{1}{r}{0.109}\\
\multicolumn{1}{r}{} & \multicolumn{1}{r}{0} & \multicolumn{1}{r}{0.02} &
\multicolumn{1}{r}{1400} & \multicolumn{1}{r}{637} & \multicolumn{1}{r}{0.174}
& \multicolumn{1}{r}{0.082}\\
\multicolumn{1}{r}{} & \multicolumn{1}{r}{0} & \multicolumn{1}{r}{0.05} &
\multicolumn{1}{r}{2296} & \multicolumn{1}{r}{655} & \multicolumn{1}{r}{0.186}
& \multicolumn{1}{r}{0.087}\\\hline
\multicolumn{1}{r}{0.25} & \multicolumn{1}{r}{0} & \multicolumn{1}{r}{0} &
\multicolumn{1}{r}{778} & \multicolumn{1}{r}{1000} & \multicolumn{1}{r}{0.182}
& \multicolumn{1}{r}{0.267}\\
\multicolumn{1}{r}{} & \multicolumn{1}{r}{0.05} & \multicolumn{1}{r}{0} &
\multicolumn{1}{r}{ 2562} & \multicolumn{1}{r}{1057} &
\multicolumn{1}{r}{0.667} & \multicolumn{1}{r}{0.275}\\
\multicolumn{1}{r}{} & \multicolumn{1}{r}{0.1} & \multicolumn{1}{r}{0} &
\multicolumn{1}{r}{3651} & \multicolumn{1}{r}{1078} &
\multicolumn{1}{r}{0.937} & \multicolumn{1}{r}{0.283}\\
\multicolumn{1}{r}{} & \multicolumn{1}{r}{0} & \multicolumn{1}{r}{0.02} &
\multicolumn{1}{r}{1532} & \multicolumn{1}{r}{1066} &
\multicolumn{1}{r}{0.193} & \multicolumn{1}{r}{0.276}\\
\multicolumn{1}{r}{} & \multicolumn{1}{r}{0} & \multicolumn{1}{r}{0.05} &
\multicolumn{1}{r}{2556} & \multicolumn{1}{r}{1182} &
\multicolumn{1}{r}{0.230} & \multicolumn{1}{r}{0.278}\\\hline
\end{tabular}
\caption{Simulation with LRS scenario: maximum MAEs and mean angles for  $p=50$ and  $n=100$}\label{TabLRSCepiM}%
\end{table}%

\end{center}

It is seen that when $d=0.5$ MM is unaffected by contamination and again
outperforms PACE in all cases. When $d=0.25$ MM has a lower MAE for
$\varepsilon_{\mathrm{case}}>0$. When $\varepsilon_{\mathrm{case}}=0.$ PACE
has lower MAE and mean Angle than MM when $\varepsilon_{\mathrm{cell}}=0,$ and
a lower maximum mean Angle when $\varepsilon_{\mathrm{cell}}>0.$ Since PACE
makes a clever use of the information when data is sparse, this result is more
on line with what would be expected.

It can be concluded that MM is resistant to both case- and cell-contamination,
that its efficiency with respect to PACE is high for decimation rate $d=0.5,$
while it may depend on the data structure for $d=0.25.$

\section{Computing algorithm of the MM-estimator\label{SecCompMM}}

The computation proceeds one component at a time. Define for $i=1,..,n$ and
$j=1,...,p$
\[
J_{i}=\{j:\ x_{ij}~\text{is non-missing}\},\ \text{\ }I_{j}=\{i:\ x_{ij}%
~\text{is non-missing}\}.
\]

At the beginning (\textquotedblleft zero components\textquotedblright): apply
robust nonparametric regression to obtain robust and smooth local location and
scale values $\widehat{\mu}_{0j}$ and $\widehat{\sigma}_{0j},$ $j=1,...,p$ as
follows. Let $S\left(  .,.\right)  $ be a robust smoother. Let $m_{j}$ be a
location M-estimator of $\{x_{ij},~i\in I_{j}\};$ then the set $\{\widehat
{\mu}_{0j},~j=1,..,p\}$ is obtained by applying $S$ to $\left(  \tau_{j}%
,m_{j},j=1,...,p\right)  .$ Let $s_{j}$ be a $\tau$-scale of $\{x_{ij}%
-\widehat{\mu}_{0j},i\in I_{j}\}$; then the set $\{\widehat{\sigma}%
_{0j},j=1,,,,p\}$ is obtained by applying $S$ to $\left(  \tau_{j}%
,s_{j},j=1,..,p\right)  .$ The chosen smoother was the robust version of Loess
(Cleveland 1979) with a span of 0.3.

Let $y_{ij}^{\left(  0\right)  }=x_{ij}-\widehat{\mu}_{0j}.$ Compute the
\textquotedblleft unexplained variance\textquotedblright%
\[
V_{0}=\frac{1}{N}\sum_{i=1}^{n}\sum_{j\in I_{j}}^{{}}\widehat{\sigma}_{0}%
{}_{j}^{2}\rho\left(  \frac{y_{ij}^{\left(  0\right)  }}{\widehat{\sigma}%
_{0}{}_{j}}\right)  ~~\mathrm{with~}\ N=\sum_{j=1}^{p}\mathrm{card}\left(
I_{j}\right)  .
\]

For component 1 use the $y_{ij}^{\left(  0\right)  }$ as input and compute%

\begin{equation}
\left(  \widehat{\boldsymbol{a}}^{\left(  1\right)  },\widehat
{\boldsymbol{\beta}}^{\left(  1\right)  },\boldsymbol{\mu}^{\left(  1\right)
}\right)  =\arg\min_{\mathbf{\alpha,\beta,}\boldsymbol{\mu}}\sum_{i=1}^{n}%
\sum_{j\in I_{j}}\widehat{\sigma}_{0j}^{2}\rho\left(  \frac{y_{ij}^{\left(
0\right)  }-\widehat{y}_{ij}^{\left(  0\right)  }\left(  \boldsymbol{\alpha
,\beta,\mu}\right)  }{\widehat{\sigma}_{0j}}\right)  . \label{minComp1}%
\end{equation}

The minimum is computed iteratively, starting from a deterministic initial
estimator to be described in Section \ref{secInitial}.

Compute the residuals $y_{ij}^{\left(  1\right)  }=y_{ij}^{\left(  0\right)
}-\widehat{y}_{ij}^{\left(  0\right)  }\left(  \widehat{\boldsymbol{\alpha}%
}\boldsymbol{,}\widehat{\boldsymbol{\beta}}\boldsymbol{,}\widehat
{\boldsymbol{\mu}}\right)  .$ Apply a smoother to compute local residual
scales $\widehat{\sigma}_{1}{}_{j}$ and the \textquotedblleft unexplained
variance\textquotedblright\ with one component:%
\[
V_{1}=\frac{1}{N}\sum_{i=1}^{n}\sum_{j\in I_{j}}\widehat{\sigma}_{1}{}_{j}%
^{2}\rho\left(  \frac{y_{ij}^{\left(  1\right)  }}{\widehat{\sigma}_{1}{}_{j}%
}\right)  .
\]
For component $k$ we have%
\begin{equation}
\left(  \widehat{\boldsymbol{a}}^{\left(  k\right)  },\widehat
{\boldsymbol{\beta}}^{\left(  k\right)  },\boldsymbol{\mu}^{\left(  k\right)
}\right)  =\arg\min_{\mathbf{\alpha,\beta,}\boldsymbol{\mu}}\sum_{i=1}^{n}%
\sum_{j\in I_{j}}\widehat{\sigma}_{k-1,j}^{2}\rho\left(  \frac{y_{ij}^{\left(
k-1\right)  }-\widehat{y}_{ij}^{\left(  k-1\right)  }\left(
\boldsymbol{\alpha,\beta,\mu}\right)  }{\widehat{\sigma}_{k-1,j}}\right)  .
\label{minCompk}%
\end{equation}

Each component is orthogonalized with respect to the former ones. The
procedure stops either at a fixed number of components or when the proportion
of explained variance (\ref{defPropExMM}) is larger than a given value (e.g. 0.90).

\subsection{The iterative algorithm\label{secIterAlgo}}

Computing each component requires an iterative algorithm and starting values.
The algorithm is essentially one of \textquotedblleft alternating
regressions\textquotedblright.

Recall that at each step $\boldsymbol{\alpha\in}R^{m}$ and $\boldsymbol{\beta
\in}R^{n}$ are one-dimensional. Put as usual $\psi=\rho^{\prime}$ and
$W(s)=\psi\left(  s\right)  /s.$ Put for brevity $h\left(  t\right)
=\sum_{l=1}^{m}\alpha_{l}B_{l}\left(  t\right)  $ where $B_{i}$ are the
elements of the spline basis.

Differentiating the criterion in (\ref{minCompk}) yields a set of estimating
equations that can be written in fixed-point form, yielding a
\textquotedblleft weighted alternating regressions\textquotedblright\ scheme.
To simplify the notation the superscript $\left(  k-1\right)  $ will be
dropped from $y_{ij}^{\left(  k-1\right)  }$ and $\widehat{y}_{ij}^{\left(
k-1\right)  }.$ Put
\[
w_{ij}=w_{ij}\left(  \boldsymbol{\alpha,\beta,\mu}\right)  =W\left(
\frac{y_{ij}-\widehat{y}_{ij}\left(  \boldsymbol{\alpha,\beta,\mu}\right)
}{\widehat{\sigma}_{k-1,j}}\right)  .
\]
Then $\mu_{j}$ and $\beta_{i}$ can be expressed as weighted residual means and
weighted univariate least squares regressions, respectively:%
\[
\mu_{j}=\frac{1}{\sum_{i\in I_{j}}w_{ij}}\sum_{i\in I_{j}}w_{ij}\left(
y_{ij}-\beta_{i}h\left(  t_{j}\right)  \right)  ,
\]%
\[
\beta_{i}=\frac{\sum_{j\in J_{i}}w_{ij}h\left(  t_{j}\right)  \mathbf{(}%
y_{ij}-\mu_{j})}{\sum_{j\in J_{i}}w_{ij}h\left(  t_{j}\right)  ^{2}}%
\]
and $\boldsymbol{\alpha}$ is the solution of%
\[
\sum_{i=1}^{n}\sum_{j\in J_{i}}w_{ij}\left(  y_{ij}-\mu_{j}\right)  \beta
_{i}\mathbf{b}\left(  t_{j}\right)  =\sum_{i=1}^{n}\sum_{j\in J_{i}}%
w_{ij}\beta_{i}^{2}\mathbf{b}\left(  t_{j}\right)  \mathbf{b}\left(
t_{j}\right)  ^{\prime}\boldsymbol{\alpha}%
\]
with $\mathbf{b}\left(  t\right)  =\left(  B_{1}\left(  t\right)
,...,B_{l}\left(  t\right)  \right)  $'.

At each iteration the $w_{ij}$ are updated. It can be shown that the criterion
descends at each iteration.

\subsection{The initial values\label{secInitial}}

For each component, initial values for $\boldsymbol{\alpha}$ and
$\boldsymbol{\beta}$ are needed. They should be deterministic, since
subsampling would make the procedure impractically slow.

\subsubsection{The initial $\boldsymbol{\alpha}$}

For $k,l\in\left\{  1,...,p\right\}  $ call $N_{kl}$ the number of cases that
have values in both $t_{k}$ and $t_{l}:$%
\[
N_{kl}=\#\left(  I_{k}\cap I_{l}\right)  .
\]

In longitudinal studies, many $N_{kl}$ may null or very small.

Compute a (possibly incomplete) $p\times p$ matrix $\boldsymbol{\Sigma
=[}\sigma_{kl}\boldsymbol{]}$ of pairwise robust covariances of $\left(
y_{ik},y_{il}:i\in I_{k}\cap I_{l}\right)  $ with the Gnanadesikan-Kettenring
procedure:%
\[
\mathrm{Cov}\left(  X,Y\right)  =\frac{1}{4}\left(  S\left(  X+Y\right)
^{2}-S\left(  X-Y\right)  ^{2}\right)  ,
\]
where $S$ is a robust dispersion. Preliminary simulations led to the choice of
the $Q_{n}$ estimator of Rousseeuw and Croux (1993).

More precisely; for $N_{kl}\geq3$ compute $\sigma_{kl}$ as above. If
$\min_{kl}N_{kl}$ is \textquotedblleft large enough\textquotedblright\ (here:
$\geq10)$ use the resulting $\mathbf{\Sigma.}$

Otherwise apply a two-dimensional smoother to improve $\mathbf{\Sigma}$ and to
fill in the missing values. The bivariate Loess was employed for this purpose.

Then compute the first eigenvector $\mathbf{e}$ of $\boldsymbol{\Sigma}$ (note
that $\boldsymbol{\Sigma}$ is not guaranteed to be positive definite and hence
further principal components may be unreliable). \medskip Given $\mathbf{e,}$
smooth it using the spline basis. This $\mathbf{\alpha}$ follows from
(\ref{defEmatrix}).

\subsubsection{The initial $\boldsymbol{\beta}$}

For $i=1,...,n$ the initial $\beta_{i}$ is a robust univariate regression of
$y_{ij}$ on $h\left(  t_{j}\right)  $ $(j\in J_{i}),$ namely the $L_{1}$
regression, which is fast and reliable.

Note that only cellwise outliers matter at this step.

\subsection{The final adjustment\label{SecFinal}}

Note that the former steps yield only an approximate solution of
(\ref{defineMM}) since the components are computed one at a time. In order to
improve the approximation a natural procedure is as follows. After computing
$q$ components we have a $p\times q$-matrix $\mathbf{U}$ of principal
directions with elements
\[
u_{jk}=\sum_{l=1}^{m}B_{l}\left(  t_{j}\right)  \alpha_{kl},
\]
an $n\times p$-matrix of weights $\mathbf{W.}$ and a location vector
$\mathbf{\mu.}$ Then a natural improvement is --keeping $\mathbf{U}$ and
$\mathbf{\mu}$ fixed-- to recompute the $\beta$s by means of univariate
weighted regressions with weights $w_{ij}.$ Let $\mathbf{\beta}_{i.}%
=[\beta_{ik},k=1,..,p],$ and set%
\[
\mathbf{\beta}_{i}=\arg\min_{\mathbf{\beta\in}R^{q}}\sum_{j\in J_{i}}%
w_{ij}\left(  x_{ij}-\mu_{j}-\mathbf{U\beta}\right)  ^{2}.
\]
The effect of this step in the case of complete data is negligible, but it
does improve the estimator's behavior for incomplete data.

However, it was found out that the improvement is not good enough when the
data are very sparse. For this reason another approach was used, namely, to
compute $\mathbf{\beta}_{i}$ as a regression M-estimate. Let $\mathbf{z}%
_{i}=(x_{ij}-\mu_{j}:j\in J_{i})$ and $\mathbf{V=[}v_{jk}\mathbf{]}$ with
$v_{jk}=u_{jk}$ for $j\in J_{i}.$ Then $\mathbf{\beta}_{i}$ is a bisquare
regression estimate of $\mathbf{z}_{i}$ on $\mathbf{V,}$ with tuning constant
equal to 4, using $L_{1}$ as a starting estimate. Note that here only cell
outliers matter, and therefore $L_{1}$ yields reliable starting values. The
estimator resulting from this step does not necessarily coincide with
(\ref{defineMM}), but simulations show that it is much better than the
\textquotedblleft natural\textquotedblright\ adjustment described above when
the data are very sparse.

\section{The \textquotedblleft naive\textquotedblright\ estimator:
details\label{secNaifDetails}}

In step 1 of Section \ref{secDefNaive}, compute for each $\mathbf{x}_{i}$
robust local location and scatter estimates $\widetilde{\mu}_{i}%
,\widetilde{\sigma}_{i}$. The \textquotedblleft cleaned\textquotedblright%
\ values are
\[
\widetilde{x}_{ij}=\widetilde{\mu}_{i}+\widetilde{\sigma}_{i}\psi\left(
\frac{x_{ij}-\widetilde{\mu}_{i}}{\widetilde{\sigma}_{i}}\right)  ,
\]
where $\psi$ is the bisquare $\psi$-function with tuning constant equal to 4.

The ordinary robust PCs of step 2 are computed using the cleaned data
$\widetilde{x}_{ij}$ with the S-M estimator of (Maronna 2005). Call
$\{\widehat{\mathbf{x}}_{i}^{\left(  q\right)  }\}$ the fit for $q$ components
and put $r_{i}^{\left(  q\right)  }=$ $\left\Vert \mathbf{x}_{i}%
-\widehat{\mathbf{x}}_{i}^{\left(  q\right)  }\right\Vert ,$ $i=1,...,n.$ Then
the estimator minimizes $S\left(  r_{i}^{\left(  q\right)  }%
,\ \ i=1,..,n\right)  $ where $S$ is the bisquare M-scale. The
\textquotedblleft proportion of unexplained variance\textquotedblright\ is
\[
\frac{S\left(  r_{i}^{\left(  q\right)  },\ \ i=1,..,n\right)  }{S\left(
r_{i}^{\left(  0\right)  },\ \ i=1,..,n\right)  }.
\]

The number of knots in step 3 is chosen through generalized cross-validation.

\section{References}

Bali, J.L., Boente, G., Tyler, D.E. and Wang, J-L. (2011). Robust functional
principal components: a projection-pursuit approach. \emph{The Annals of
Statistics,} \textbf{39}, 2852--2882.

Bay, S. D. (1999), The UCI KDD Archive [http://kdd.ics.uci.edu], University of
California, Irvine, Dept. of Information and Computer Science.

Boente, G. and Salibian-Barrera, M. (2015). S-Estimators for Functional
Principal Component Analysis. \emph{Journal of the American Statistical
Association, }\textbf{110, }1100-1111.

Cevallos Valdiviezo, H. (2016). On Methods for Prediction Based on Complex
Data with Missing Values and Robust Principal Component Analysis, PhD thesis,
Ghent University (supervisors Van Aelst S. and Van den Poel, D.).

Cleveland, W.S. (1979). Robust Locally Weighted Regression and Smoothing
Scatterplots. \emph{Journal of the American Statistical Association,}
\textbf{74,} 829-836.

James, G., Hastie, T.G., and Sugar, C.A. (2001). Principal Component Models
for Sparse Functional Data. \emph{Biometrika}, \textbf{87}, 587-602.

Lee, S., Shin, H. and Billor, N. (2013). M-type smoothing spline estimators
for principal functions. \emph{Computational Statistics and Data Analysis},
\textbf{66, }89-100.

Maronna, R. (2005). Principal components and orthogonal regression based on
robust scales. \emph{Technometrics}, \textbf{47}%
\c{}
264-273.\newline

Maronna, R.A., Martin, R.D. and Yohai, V.J. (2006). \emph{Robust
Statistics:\ Theory and Methods. }John Wiley and Sons, Chichester.

Rousseeuw, P.J. and Croux, C. (1993). Alternatives to the Median Absolute
Deviation. \emph{Journal of the American Statistical Association,}
\textbf{88}, 1273--1283.

Yao, F., M\"{u}ller, H-G. and Wang, J-L. (2005). Functional Data Analysis for
Sparse Longitudinal Data. \emph{Journal of the American Statistical
Association, }\textbf{100, }577-590.

Yohai, V.J. (1987). High Breakdown-Point and High Efficiency Robust Estimates
for Regression. \emph{The Annals of Statistics,} \textbf{15, 642-656.}

\end{document}